\documentclass[%
 reprint,
 amsmath,amssymb,
 aps,
prr,
floatfix,
]{revtex4-2}
\usepackage{graphicx}
\usepackage{dcolumn}
\usepackage{bm}
\usepackage{color}

\newcommand{\bk}{\mathbf{k}}
\newcommand{\bkp}{{\mathbf{k}'}}
\newcommand{\bq}{\mathbf{q}}

\begin{document}


\title{On the effect of disorder in a $p$-wave flat-band superconductor}

\author{Michele Governale}

 \email{michele.governale@vuw.ac.nz}
\affiliation{%
School of Chemical and Physical Sciences and MacDiarmid Institute for Advanced Materials and Nanotechnology,\\
Victoria University of Wellington, Wellington 6140\\
 New Zealand}%

\date{\today}

\begin{abstract}
We present a theoretical study of flat-band superconductivity for fully spin-polarized triplet pairing with $p$-wave symmetry. We consider the effect of disorder and calculate the disorder-averaged Green's functions diagrammatically in first-order Born approximation. In the clean limit, we find that, similarly to the case of $s$-wave pairing in a flat-band, both the gap and the critical temperature depend linearly on the attractive interaction strength. We derive the critical-temperature suppression formula for the flat-band case and find that $p$-wave flat-band superconductivity is more robust than the standard case of a band with finite density of states, particularly in more disordered samples. 
\end{abstract}

\maketitle


\textit{Introduction.} 
In a conventional superconductor, both the superconducting gap $|\Delta|$ and the critical temperature $T_{C}$ depend on the attractive interaction strength $V$ as $\propto e^{-1/(V N(E_F) \Omega)}$, where $N(E_F)$ is the density of states (DOS) at the Fermi energy and $\Omega$ the volume of the sample \cite{Tinkham}. Usually, the week coupling limit $V N(E_F) \Omega\lessapprox 1$ is applicable \cite{Tinkham} and  this leads to low-values of the critical temperature. The situation is different when superconductivity occurs in a flat-band \cite{Khodel1990} as the DOS diverges and both the gap and the critical temperature depend linearly on the interaction strength $V$. Thus flat-band superconductivity (FBS) provides a mechanism to obtain  much higher critical temperatures for equal strengths of the attractive interaction.  

FBS has been proposed as a mechanism to achieve high $T_C$ in carbon-based materials \cite{Volovik2018}, such as  graphite with rhombohedral stacking \cite{Kopnin2011,Kopnin2013,Lothman2017}, twisted bylayer graphene \cite{shaginyan2021} and strained graphene \cite{Kauppila2016}. Experimental signatures of  FBS  have been found in bilayer graphene \cite{Cao2018}, where a zero-resistance  state is observed when the angle between the two graphene sheets in the twisted bilayer  is equal to certain \textit{magic} values at which the single-particle spectrum exhibits a flat band near the Fermi energy. FBS has also been considered as the mechanism leading to the observation of high-$T_C$ in highly oriented pyrolytic graphite \cite{Ballestar2013,Esquinazi2014,Precker2016}. Other systems where FBS could play a significant role include strongly correlated materials \cite{Khodel1990,Shaginyan2010,shaginyan2021} and interfaces of topological II-VI semiconductors \cite{Tang2014}. 
So far FBS has been investigated in the context of spin-singlet pairing. 

In this Letter we consider FBS in the case of triplet superconductivity. Triplet superconductors \cite{Mackenzie2003,Mineev2017,Saxena2000,Aoki2001,Pfleiderer2001,Pfleiderer2009} have recently attracted considerable interest  for their potential applications in the field of superconducting spintronics \cite{Lindner2015,Eschrig2015}. 
In contrast to singlet pairing, triplet pairing is very sensitive to the presence of disorder \cite{Mineev99} which in a system with a finite DOS leads to a suppression of the critical temperature described by 
\begin{align}
\label{eq:TCsuppression-std}
  \log\left(\frac{T_{C0}}{T_C}\right)=
  \psi\left(\frac{1}{2}+\frac{1}{\pi 4 k_B T_c\tau} \right)-\psi\left(\frac{1}{2}\right),
  \end{align}
where $T_{C0}$ is the critical temperature in the clean limit, $\tau$ the quasiparticle scattering time, $\psi$ the digamma function, and $k_B$ the Boltzmann constant. Here and in the following, we set $\hbar=1$. 

In order to establish whether the presence of a flat-band provides a viable mechanism to establish triplet superconductivity with an experimentally-accessible critical temperature, it is necessary to study the role of disorder in triplet FBS. This is the main aim of the present Letter. 

The experimental motivation for our study is provided by the discovery of superconductivity in  samarium nitride (SmN) below 4 K and of the coexistence of superconductivity and ferromagnetism in this rare-earth compound \cite{Anton2016}. While no direct measure of the symmetry of the gap is available, the presence of a large exchange splitting in the conduction band strongly indicates equal-spin triplet pairing. However, how the $p$-wave triplet correlations can survive in the presence of disorder is still an open question. SmN is the only member of the rare-earth nitrides in which a superconductive transition has been observed. This could be due to the presence in SmN of a flat $f$ band near the Fermi energy \cite{Holmes-Hewett2019}. Triplet FBS could be the mechanism yielding superconductivity in the semiconductor SmN. 

\begin{figure}
    \centering
    \includegraphics[width=0.8\columnwidth]{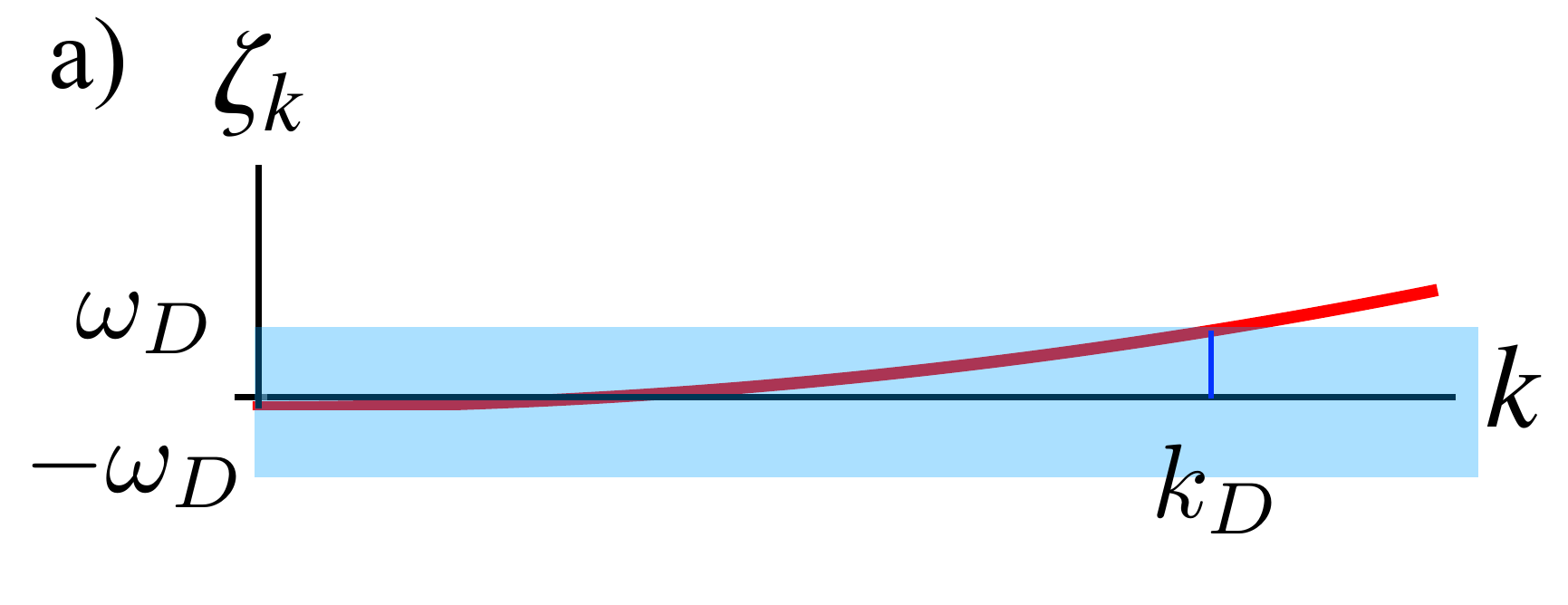}\\
    \includegraphics[width=0.8\columnwidth]{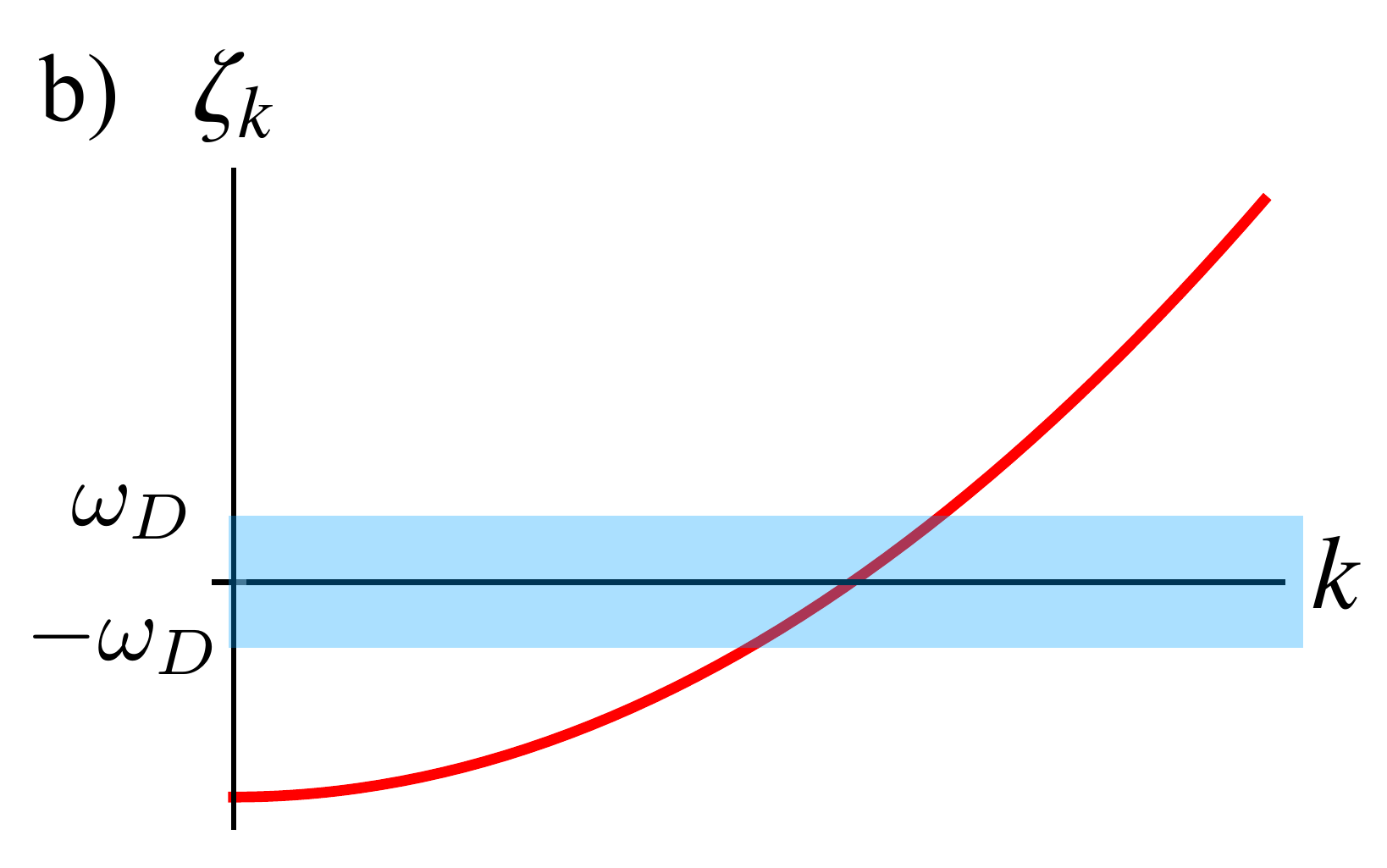}
    \caption{Schematic description of the band structure for: a) the case of a flat band at the Fermi energy with $E_F\ll \omega_D$; b) the case of a parabolic band with $E_F\gg \omega_D$. The attractive interaction is present for states in an energy window of width $2\omega_D$ around the Fermi energy (illustrated by shaded rectangles in the plots).}
    \label{fig:bands}
\end{figure}

\textit{Model \& Formalism.}
We consider a fully spin-polarised triplet superconductor described by the following mean-field Hamiltonian
\begin{align}
\label{eq:H0}
H_0=\sum_{\bk}\zeta_{\bk}c_{\bk}^\dagger c_{\bk}
-\frac{1}{2}\sum_{\bk}\left( 
\Delta(\bk)c_{\bk}^\dagger c_{-\bk}^\dagger+
\Delta(\bk)^{*}c_{-\bk} c_{\bk}
\right)
\end{align}
where the operator $c_{\bk}^{(\dagger)}$ annihilates(creates) an electron with momentum $\bk$. We suppress the spin index as only majority spins are present in the system. We denote the single-particle excitation energy by $\zeta_{\bk}$. 
We consider the case of a flat band at the Fermi energy, that is $\zeta_{\bk}\rightarrow 0$. We can imagine the flat band as the limit of a parabolic band with diverging effective mass: $\zeta_{\bk}=\alpha (\bk^2-k_F^2)$ for $\alpha\rightarrow 0$. The attractive interaction is present in a energy window of width $2\omega_D$ around the Fermi energy. 
As shown in Fig.~\ref{fig:bands}(a), in the case of a flat band the attractive interaction will be present for states with $|\mathbf{k}|\le k_D$. We will contrast FBS with the standard case in which we have a band with finite DOS and $E_F\gg 2\omega_D$. In this latter case, we can approximate $\zeta_{\bk}\approx v_F (|\bk|-k_F)$ in the energy window relevant for superconductivity. The standard case is depicted in Fig.~\ref{fig:bands}(b).

\begin{figure}
    \centering
    \raisebox{0.2\columnwidth}{a)}~ \includegraphics[width=0.4\columnwidth]{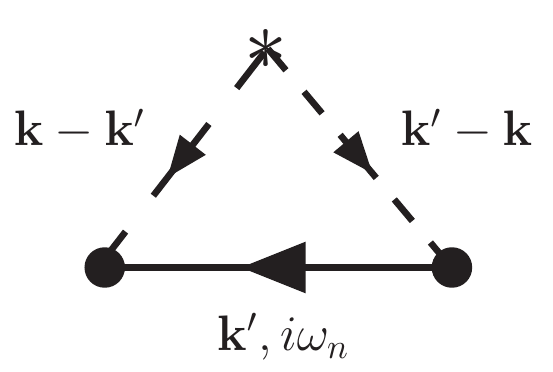}\hspace{0.05\columnwidth} \raisebox{0.2\columnwidth}{b)}~
    \includegraphics[width=0.4\columnwidth]{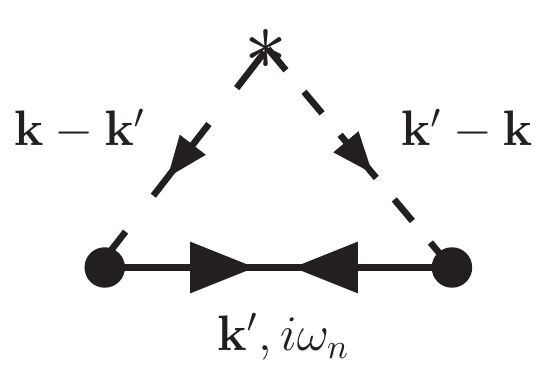}
    \caption{Examples of diagrams contributing to the disorder selfenergy in first-order Born approximation. The graphical symbols in the diagrams have the following meaning: 
    i) solid line with one arrow and label $\bkp$ denotes $G^0(\bkp,i\omega_n)$; ii) the solid line with two arrows  and label $\bkp$ denotes $F^0(\bkp,i\omega_n)$;(iii) the dashed scattering line with label $\mathbf{q}$ is equal to the amplitude $u_{\mathbf{q}}$;(iv) the $\star$ indicates a momentum conserving impurity-averaged factor $n_{\text{imp}}$. The diagrams correspond to: 
    a) $\Sigma_{ee}(\bk,i\omega_n)$; b) $-\Sigma_{he}(\bk,i\omega_n)$.}
    \label{fig:selfenergy}
\end{figure}

We model the disorder by $N_{\text{imp}}$ impurities located at positions $\mathbf{P}_j$ and average over the positions of the impurities. 
The impurity potential is given by 
\begin{align}
U=\frac{1}{\Omega}\sum_{j=1}^{N_{\text{imp}}}\sum_{\bk,\bq}u_{\bq} e^{-i \bq\cdot \mathbf{P}_j}c_{\bk}^\dagger c_{\bk-\bq},
\end{align}
where $u_{\bq}$ is the Fourier transform of the potential $u(\mathbf{r})$ of a single impurity located at the origin. 
We use the Nambu representation and define the operators  
$\mathbf{\Psi}_\bk  =\left(c_\bk, c_{-\bk}^{\dagger}\right)^{\text{T}}$ and 
$\mathbf{\Psi}_{\bk}^{\dagger}  =\left(c_{\bk}^{\dagger}, c_{-\bk}\right)$, where T denotes the transpose. 
We define the full Matsubara Green's function after impurity averaging as \footnote{After averaging over the impurity configurations, momentum is conserved since translational invariance is restored.} 
\begin{align}
 \mathbf{G}_{\bk}(i\omega_n)=-\int_0^{\beta}d\tau e^{i \omega_n \tau}\overline{\langle T_\tau \mathbf{\Psi}_\bk (\tau) \mathbf{\Psi}_\bk^\dagger (0)\rangle}, 
\end{align}
where $T_\tau$ is the time-ordering operator in imaginary time, $\omega_n$ are the Fermionic Matsubara frequencies, $\beta=1/(k_B T)$, and the over line indicates averaging over the positions of the impurities. 
We write the Green's function as a matrix
\begin{align}
    \mathbf{G}_{\bk}(i\omega_n)=
    \left(\begin{array}{cc}
     G(\bk, i\omega_n) & -F(-\bk, i\omega_n)^*\\
     F(\bk, i\omega_n) &- G(-\bk, i\omega_n)^*
    \end{array}\right),
\end{align}
where we have made use of the fact that there are only two independent matrix elements.
The full Green's function after impurity averaging obeys the Dyson equation
\begin{align}
\label{eq:Dyson}
    \mathbf{G}_{\bk}(i\omega_n)=\mathbf{G}^0_{\bk}(i\omega_n)+\mathbf{G}^0_{\bk}(i\omega_n)\mathbf{\Sigma}_\bk(i\omega_n) \mathbf{G}_{\bk}(i\omega_n),
\end{align}
where $\mathbf{\Sigma}_\bk(i\omega_n)$ is the self-energy due to the impurities and $\mathbf{G}^0_{\bk}(i\omega_n)$ is the Green's function in the absence of disorder. The free Green's functions can be computed, for example, by means of the equation-of-motion method \cite{Bruus}. The calculation of the free Green's function is not complicated but tedious and here we will simply report the result: $
G^0(\bk,i\omega_n)=(i\omega_n+\zeta_\bk)/[(i\omega_n)^2-E_\bk^2]$ and $
F^0(\bk,i\omega_n)=-(\Delta_\bk^*)/[(i\omega_n)^2-E_\bk^2]$, 
where we have defined the excitation energies of the superconductor as 
$E_\bk=\sqrt{\zeta_\bk^2+|\Delta(\bk)|^2}$.
We compute the disorder self-energy  $\Sigma_\bk(i\omega_n)$ in first-order Born approximation by means of a standard diagrammatic technique to perform the average over the impurity configurations \cite{Bruus, Abrikosov}. This yields
\begin{subequations}
\begin{align}  
\nonumber
   \Sigma_{ee}(\bk,i\omega_n)&=n_{\text{imp}}\frac{1}{\Omega}\sum_{\bkp}|u_{\bkp-\bk}|^2 G^0(\bkp,i\omega_n)\\ &\approx -i \text{sign}(\omega_n)\frac{1}{2 \tau_\bk}\\
   \nonumber
   \Sigma_{hh}(\bk,i\omega_n)&=-n_{\text{imp}}\frac{1}{\Omega}\sum_{\bkp}|u_{\bkp-\bk}|^2 G^0(-\bkp,i\omega_n)^*\\ &\approx -i \text{sign}(\omega_n)\frac{1}{2 \tau_{\bk}}\\
   \label{eq:she}
   \Sigma_{he}(\bk,i\omega_n)&=-n_{\text{imp}}\frac{1}{\Omega}\sum_{\bkp}|u_{\bkp-\bk}|^2 F^0(\bkp,i\omega_n)\approx 0\\
   \label{eq:seh}
   \Sigma_{he}(\bk,i\omega_n)&=n_{\text{imp}}\frac{1}{\Omega}\sum_{\bkp}|u_{\bkp-\bk}|^2 F^0(-\bkp,i\omega_n)^*\approx 0  ,
\end{align}
\end{subequations}
where $n_{\text{imp}}=N_{\text{imp}}/\Omega$ is the impurity density.
The expression for the scattering rate $1/\tau_\bk$ is given in the Appendix. When solving for the critical temperature, $\Delta(\bk)\rightarrow 0$ and the scattering rate reduces to the one for the normal state, that is
$1/{\tau_\bk}=2\pi n_{\text{imp}}\frac{1}{\Omega}\sum_{\bkp}|u_{\bkp-\bk}|^2 \delta(\zeta_\bk-\zeta_\bkp)$.  For the standard case, we assume both the DOS and $|u_{\bkp-\bk}|^2\approx |u_0|^2$ to be constant, then the scattering rate is independent of $\bk$ and equal to  $1/{\tau}=2 \pi n_{\text{imp}} |u_0|^2 N(E_F)$. 
For the case of a flat band, we assume some weak residual dispersion of the band, so that the DOS $N_{\text{fb}}$ is large but not infinite and the 
scattering rate $1/\tau$ remains finite.  The off-diagonal elements of the selfenergy vanish since $\Delta(\bkp)=-\Delta(-\bkp)$ and $|u_{\bkp-\bk}|$ varies little as a function of $\bkp$. 
We show an example of the diagrams contributing to the selfenergy in Fig.~\ref{fig:selfenergy}.

The full Green's function is obtained by solving the Dyson Eq.~(\ref{eq:Dyson}). 
In particular we find
\begin{align}
\label{eq:G_final}
G(\bk,i\omega_n)&=\frac{i\omega_n +  i\text{sign}(\omega_n)\frac{1}{2\tau_\bk}+\zeta_\bk}
{\left( i\omega_n +  i\text{sign}(\omega_n)\frac{1}{2\tau_{\bk}} \right)^2-E_\bk^2 }\\
\label{eq:F_final}
F(\bk,i\omega_n)&=-\frac{\Delta(\bk)^*}
{\left( i\omega_n +  i\text{sign}(\omega_n)\frac{1}{2\tau_{\bk}} \right)^2-E_\bk^2 }.
\end{align}

\textit{The gap equation.}
The gap can be expressed in terms of the anomalous Matsubara Green's function as 
\begin{align*}
 \Delta(\bk)&=\sum_{\bkp}V_2\, \bk\cdot\bkp\langle c_{-\bkp}c_{\bkp}\rangle
 \\
  &=\sum_{\bkp}V_2\, \bk\cdot\bkp\left(\frac{1}{\beta}\sum_{\omega_n}e^{i \omega_n 0^-} F(\bkp,i\omega_n)\right)^*
\end{align*}
where $V_2 \bk\cdot\bkp$ is the interaction strength for $p$-wave pairing. 
Performing the sum over the Matsubara frequencies yields
\begin{align}
\nonumber
    \Delta(\bk)&=\sum_{\bkp}V_2\, \bk\cdot\bkp\frac{\Delta(\bkp)}{2E_{\bkp}}\\
    \label{eq:gap_disorder}
    &\int d\omega f(\omega)\left[   L_{1/\tau_{\bkp}}(\omega+E_{\bkp})-L_{1/\tau_{\bkp}}(\omega-E_{\bkp})\right],
\end{align}
where $L_\Gamma(x)$ is the Lorentzian curve of width $\Gamma$ defined as
 $ L_\Gamma(x)=\frac{1}{\pi}\frac{\Gamma/2}{x^2+(\Gamma/2)^2}$. 
The gap equation (\ref{eq:gap_disorder}) fully describes the effect of disorder in a triplet superconductor. 
The integral in $d\omega$ can be performed analytically
and the gap-equation becomes 
\begin{align}
\label{eq:gap_disorder2}
   \Delta(\bk)=\sum_{\bkp}V_2\, \bk\cdot\bkp\, \frac{\Delta(\bkp)}{2E_{\bkp}} \frac{2}{\pi}\text{Im}\left\{
    \psi\left[\frac{1}{2}+ \frac{\beta}{ 4\pi\tau_{\bkp}}+ i\frac{\beta}{2\pi}E_{\bkp}\right]\right\}. 
\end{align}
In the limit of vanishing disorder, Eq.~(\ref{eq:gap_disorder2}) reduces to the known result for the clean limit, that is $
    \Delta(\bk)=\sum_{\bkp}V_2\, \bk\cdot\bkp\frac{\Delta(\bkp)}{2E_{\bkp}}
    \text{tanh}\left(\frac{\beta}{2}E_{\bkp}\right)$.

\textit{Results}
We start by reviewing the results for the standard case. If we choose the $p_z$ symmetry for the gap, that is $\Delta(\bk)=\Delta_{\text{std}}(T) \frac{k_z}{k_{F}}$, we obtain for the gap at zero temperature in the clean limit 
\begin{align}
  \label{eq:delta-standard}
      \Delta_{\text{std}}^{(0)}(0)=2 e^{1/3}\omega_D\exp\left( -\frac{3}{\lambda N(E_F)}\right),      
  \end{align}
  where the superscript $(0)$ indicates the clean limit, $\lambda=\Omega V_2 k_F^2$ and $N(E_F)$ is the density of states at the Fermi energy $N(E_F)={k_F^2}/(2v_F \pi^2)$. The critical temperature in the clean limit is related to the gap by 
      $k_B T_{C0}=(e^{\gamma-1/3}/\pi) \Delta_{\text{std}}^{(0)}(0).$
The suppression formula of Eq.~(\ref{eq:TCsuppression-std}), can be derived from the gap equations with and without  disorder (\ref{eq:gap_disorder}) setting $T=T_C-0^+$, i.e. when $\Delta(\bk)\rightarrow 0$.

Now, we proceed to consider fully spin-polarized triplet pairing with $p$-wave symmetry in a flat band, that is we take $\zeta_\bk=0$. 
It is instructive to calculate the gap in the clean case at zero temperature. We consider the case $\Delta(\bk)=\Delta_{\text{fb}}(T) (k_z/k_D)$, but the other possible $p$-wave symmetries yield very similar results. Notice that we choose $k_D$ as the relevant scale for momentum.
Starting from the clean-limit of Eq.~(\ref{eq:gap_disorder}) and using spherical coordinates $\bkp=k'(\sin\theta\cos\phi,\sin\theta\sin\phi,\cos\theta)$, the clean-case gap equation at $T=0$ reads
\begin{align}
    1=\frac{1}{2}\frac{\Omega}{(2\pi)^2}V_2\int_0^{k_D} dk (k')^4\int_{-1}^{1}ds \frac{|s|}{\Delta_{\text{fb}}^{(0)}(0) \frac{k'}{k_D}},
\end{align}
where $s=\cos\theta$. The integrals can be performed easily and one obtains for the gap 
\begin{align}
\label{eq:delta-fb}
    \Delta_{\text{fb}}^{(0)}(0)=\frac{1}{8} \frac{\Omega}{(2 \pi)^2} V_2 k_D^5.
\end{align}
As expected for FBS, the gap is linear in the interaction strength $V_2$. We can rewrite Eq.~(\ref{eq:delta-fb}) as $ \Delta_{\text{fb}}^{(0)}(0)=(3/16) V_2 k_D^2 N_{kp}$, where $V_2 k_D^2$ is the largest possible value of the interaction energy and $N_{kp}$ is the number of $k$ points in the volume of reciprocal space where the attractive interaction is present. 
Equation~(\ref{eq:delta-fb}) needs to be contrasted to the standard case of Eq.~(\ref{eq:delta-standard}). 

We now proceed as we did before and write the gap equation at $T=T_C-0^+$ both in the clean limit and in the presence of disorder. Again we perform the flat-band limit $\zeta_\bk\rightarrow 0$. 
In the clean limit we obtain
\begin{align}
\label{eq:TC0_flat}
    1=\frac{\Omega}{(2\pi)^2}\frac{V_2}{3}\frac{k_D^5}{5}\left[\frac{1}{2 k_B T_{C0}}\right],
\end{align}
which leads to the following relation between $\Delta_{\text{fb}}^{(0)}(0)$ and $T_{C0}$:
\begin{align}
k_B T_{C0}=\frac{4}{15}\Delta_{\text{fb}}^{(0)}(0).  
\end{align}
Similarly to the case of singlet pairing \cite{Volovik2018}, FBS provides a mechanism to achieve high values of $T_C$, at least in the clean limit. The final question that will be addressed concerns whether disorder might limit the enhancement of $T_{C0}$ provided by the FBS mechanism. 
\begin{figure}[t]
    \centering
    \includegraphics[width=0.9\columnwidth]{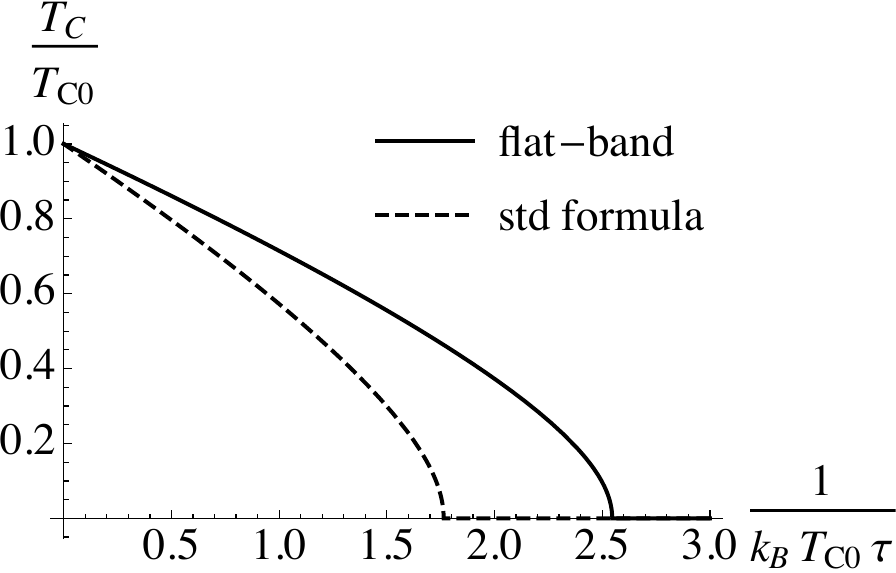}
    \caption{Comparison of the $T_C$-suppression formulae for the flat-band and the standard-case. We plot the suppression factor $T_c/T_{C0}$ as a function of the strength of disorder quantified by $1/(k_B T_{C0} \tau)$. }
    \label{fig:suppression}
\end{figure}
In the presence of disorder, the equation for $T_C$ reads
\begin{align}
\label{eq:TC_flat}
1=\frac{\Omega}{(2\pi)^2}\frac{V_2}{3}\frac{k_D^5}{5}\left(\frac{1}{2 k_B T_C}\right)\frac{2}{\pi^2}
\psi^{(1)}\left(\frac{1}{2}+\frac{1}{\pi 4 k_B T_C \tau}\right),
\end{align}
where $\psi^{(1)}$ is the trigamma function.
Combining Eqs.~(\ref{eq:TC0_flat}) and (\ref{eq:TC_flat}), we obtain the $T_C$-suppression formula for the flat-band case
\begin{align}
\label{eq:TCsuppression-flat}
\frac{T_C}{T_{C0}}=\frac{2}{\pi^2}\psi^{(1)}\left(\frac{1}{2}+\frac{1}{\pi 4 k_B T_C \tau}\right)\, .  
\end{align}
We now compare the $T_C$-suppression formula for the flat-band case, Eq.~(\ref{eq:TCsuppression-flat}), with the standard result for $p$-wave of Eq.~(\ref{eq:TCsuppression-std}). In Fig.~\ref{fig:suppression}, we plot 
$T_C/T_{C0}$  
as a function of $1/(k_B T_{C0} \tau)$, both for the flat-band case and the standard case. The expression $1/(k_B T_{C0} \tau)$ measures the strength of disorder in comparison to the critical temperature in the clean limit. For the case of FBS it is given by 
$1/(k_B T_{C0} \tau)=(3/16) (2\pi n_{\text{imp}}|u_0|^2 N_{\text{fb}})/(V_2 k_D^2 N_{kp})$, 
where $N_{\text{fb}}$ is the large DOS associated to the residual dispersion of the flat band \footnote{In order to have FBS the DOS due to the weak dispersion of the flat band needs to fulfil the relation $V_2 k_{D}^2 \Omega N_{\text{fb}}\gg 1$}. Figure \ref{fig:suppression} clearly shows that the critical temperature is less suppressed in the flat-band case in particular for larger values of $1/(k_B T_{C0} \tau)$. Therefore, FBS offers a mechanism to obtain triplet superconductivity at experimentally accessible temperatures.

\textit{Conclusions.} We have considered flat-band superconductivity for fully spin-polarised triplet pairing. Similarly to the singlet case, both the gap and the critical temperature depend linearly on the attractive interaction. We have considered the effect of impurities on the critical temperature by calculating the impurity-averaged Green's function diagrammatically in first-order Born  approximation. We have found a $T_C$ suppression formula which shows that FBS is more robust to disorder than the standard case. \\
\textit{Acknowledgements.} 
The author gratefully acknowledges extensive discussions with Ulrich Z\"ulicke  from Victoria University of Wellington. 
Joe Trodahl, Ben Ruck, and William Holmes-Hewett from the Spintronics research group at Victoria University of Wellington are gratefully acknowledged for providing experimental insight and many stimulating discussions.
\appendix*
\section{Self-energy}
In this Appendix, we consider the expression for the selfenergy
\begin{align}  
  \Sigma_{ee}(\bk,i\omega_n) 
   &=n_{\text{imp}}\frac{1}{\Omega}\sum_{\bkp}|u_{\bkp-\bk}|^2 
   \frac{i \omega_n +\zeta_\bkp}{(i \omega_n)^2-(\zeta_\bkp^2+|\Delta(\bkp)|^2)}.
\end{align}
We can write this expression as 
\begin{widetext}
\begin{align}  
  \Sigma_{ee}(\bk,i\omega_n) 
   =n_{\text{imp}}\frac{1}{\Omega}\sum_{\bkp}|u_{\bkp-\bk}|^2 
  \frac{i \omega_n +\zeta_\bkp}{2 E_\bkp}
  \left(
   \frac{1}{(i \omega_n)-E_\bkp}-\frac{1}{(i \omega_n)+E_{\bkp}}\right). 
\end{align}
\end{widetext}
In order to evaluate the selfenergy, We follow the procedure outlined in Ref.~\cite{Bruus}.
Having in mind that we will make an analytical continuation we can apply the substitution $i\omega_n\rightarrow \omega+ i \text{sign}(\omega_n)0^+$. Performing this substitution we get
\begin{widetext}
\begin{align*}  
  \Sigma_{ee}(\bk,i\omega_n)&=- 
i\pi\text{sign}(\omega_n)n_{\text{imp}}\frac{1}{\Omega}\sum_{\bkp}|u_{\bkp-\bk}|^2 
   \frac{\omega +\zeta_\bkp}{2 E_\bkp}
  \left(
   \delta(\omega-E_\bkp)-\delta(\omega+E_\bkp)\right)\\
&= - 
i\pi\text{sign}(\omega_n)n_{\text{imp}}\frac{1}{\Omega}\sum_{\bkp}|u_{\bkp-\bk}|^2 
  \left[ \frac{1}{2}
  \left(
   \delta(\omega-E_\bkp)+\delta(\omega+E_\bkp)\right)+
   \frac{\zeta_\bkp}{2 E_\bkp}\left(
   \delta(\omega-E_\bkp)-\delta(\omega+E_\bkp)\right)
   \right]
   \\  
   &\approx 
   - i\pi\text{sign}(\omega_n)n_{\text{imp}}\frac{1}{\Omega}\sum_{\bkp}|u_{\bkp-\bk}|^2 \left[
  \frac{1}{2}
  \left(
   \delta(E_\bk-E_\bkp)+\delta(E_\bk+E_\bkp)\right)+
   \frac{\zeta_\bkp}{2 E_\bkp}\left(
   \delta(E_\bk-E_\bkp)-\delta(E_\bk+E_\bkp)\right)
   \right].
\end{align*}
\end{widetext}
The last step relies on the fact that $\omega$ will be forced to be equal to $E_\bk$ by the quasiparticle spectral function. We therefor find in general that the scattering rate $1/\tau_{\bk}$ is given by the expression
\begin{widetext}
\begin{align*}  
  \frac{1}{\tau_\bk}=
   2\pi n_{\text{imp}}\frac{1}{\Omega}\sum_{\bkp}|u_{\bkp-\bk}|^2 \left[
  \frac{1}{2}
  \left(
   \delta(E_\bk-E_\bkp)+\delta(E_\bk+E_\bkp)\right)+
   \frac{\zeta_\bkp}{2 E_\bkp}\left(
   \delta(E_\bk-E_\bkp)-\delta(E_\bk+E_\bkp)\right)
   \right].
\end{align*}
\end{widetext}
When solving for the critical temperature, $\Delta(\bk)\rightarrow 0$, and we recover the normal-state scattering rate
\begin{align*}  
  \frac{1}{\tau_\bk}=2\pi n_{\text{imp}}\frac{1}{\Omega}\sum_{\bkp}|u_{\bkp-\bk}|^2 
   \delta(\zeta_\bk-\zeta_\bkp). 
\end{align*}
\bibliography{super}
\end{document}